%% file: springer.tex
\documentclass[titlepage,11pt,subeqn]{amsart}
\usepackage{amsmath,amssymb,amscd,amsxtra}
\usepackage[mathscr]{eucal}
\usepackage{palatino}
\usepackage{bigints}
\usepackage{color,colortbl}
\usepackage{latexsym}
\usepackage{mathtools}   
\usepackage{amsfonts}
\usepackage{ifthen}
\usepackage[]{graphicx}
\usepackage[]{hyperref} 
\usepackage{bbm}
\usepackage{amsthm}
\usepackage{tikz}
\usetikzlibrary{shapes,snakes}
\usetikzlibrary{arrows,decorations}
\usepackage[]{booktabs,tabularx}
\definecolor{Gray}{gray}{0.9}
\newcolumntype{b}{X}
\newcolumntype{s}{>{\hsize=.5\hsize}X}
\usepackage{empheq}
\newcommand\txtred[1]{{\color{red}#1}}

\begin{document}
\title[Codimension two defects and the Springer correspondence]{Codimension two defects and the Springer correspondence} 
 \author[Aswin Balasubramanian]{Aswin Balasubramanian}
 \address{ Theory Group, Department of Physics, University of Texas at Austin.} 
 \thanks{UTTG-27-14}
 \thanks{The author's work was made possible by the hospitality of the Theory Group, UT Austin. The author would also like to thank the Mathematical Sciences Research Institute, Berkeley for hospitality during the \textit{Introductory Workshop on Geometric Representation Theory}.}
 \begin{abstract} 
 One can associate an invariant to a large class of regular codimension two defects of the six dimensional $(0,2)$ SCFT $\mathscr{X}[j]$ using the classical Springer correspondence. Such an association  allows a simple description of S-duality of associated Gaiotto-Witten boundary conditions in $\mathcal{N}=4$ SYM for arbitrary gauge group $G$ and by extension, a determination of certain local aspects of class $\mathcal{S}$ constructions. I point out that the problem of \textit{classifying} the corresponding boundary conditions in $\mathcal{N}=4$ SYM is intimately tied to possible symmetry breaking patterns in the bulk theory. Using the Springer correspondence and the representation theory of Weyl groups, I construct a pair of functors between the class of boundary conditions in the theory in the phase with broken gauge symmetry and those in the phase with unbroken gauge symmetry. 
 \end{abstract}
\maketitle
\tableofcontents

\section{Introduction}

In recent years, the study and possible classification of defect operators in supersymmetric quantum field theories of various dimensions has been pursued with some vigor. Two major themes in this study has been connections between the study of defect operators and various ideas in geometric representation theory and the close relationship between defects in a $d$ dimensional (S)QFT and various $l$ (for $l<d$) dimensional (S)QFTs. In this note, I will discuss a particular instance where both of these themes play a prominent role. This involves a class of regular codimension two defects of the six dimensional $(0,2)$ SCFT that we will henceforth denote by $\mathscr{X}[j]$ for every Lie algebra $j$ of type $A,D,E$. On the one hand, the problem of classifying these defects turns out to be connected intimately to the classical Springer correspondence and closely related aspects of the representation theory of Weyl groups. And on the other hand, the existence of these defects in the six dimensional theory is tied to the existence of certain lower dimensional SCFTs with eight supercharges. One set of examples of the three dimensional SCFTs in question are the Gaiotto-Witten theories $T^\rho[G]$ and their mirror duals $T_{\rho^\vee}[G]$.  These theories are interesting objects by themselves and also serve as important ingredients in constructions of other four and three dimensional theories. Specifically, in four dimensions, a subset of the theories dubbed `class $\mathcal{S}$' \cite{Gaiotto:2009we,Gaiotto:2009hg} are of this type. 

In what follows, these Gaiotto-Witten theories will play an important role. In Section \ref{BCS}, their specific role in the study of boundary conditions in four dimensional $\mathcal{N}=4$ SYM and the relationship to codimension-two defects of $\mathscr{X}[j]$ is recalled. In Section \ref{springer}, the close connection of these topics to the Springer Correspondence is recalled. In Section, \ref{Classification}, the connection is exploited to provide a classification scheme that relates the existence of the Gaiotto-Witten theories to the possibility of specific symmetry breaking patters in four dimensional $\mathcal{N}=4$ SYM theory. A pair of functors between the classes of boundary conditions of the theory with gauge group $G^\vee$ and the theory with gauge group $L^\vee$ (for $L^\vee$ a suitable subgroup of $G^\vee$) play an important role in this classification scheme. While this note is largely a review of the work done in \cite{Balasubramanian:2014jca}, the narrower focus of this note allows for an elaboration of some subtler aspects like, for example, the difference between the two finite groups usually denoted by $A(O)$ and $\overline{A}(O)$ and how the difference between the two can be understood in terms of the Springer invariant.

\textbf{Acknowledgements:} I would like to thank P. Achar, D. Ben Zvi, J. Distler, A. Neitzke, N. Proudfoot for discussions.

\section{Boundary conditions for $\mathcal{N}=4$ SYM}
\label{BCS}

\begin{figure}
\input{figs/nahmhitchin.tex}
\caption{The connection between codimension two defects of $\mathscr{X}[j]$ and boundary conditions for 4d $\mathcal{N}=4$ SYM. The Nahm ($\rho$) and Hitchin ($\rho^\vee$) labels associated to a regular codimension two defect are highlighted.}
\label{nahmhitchin}
\end{figure}
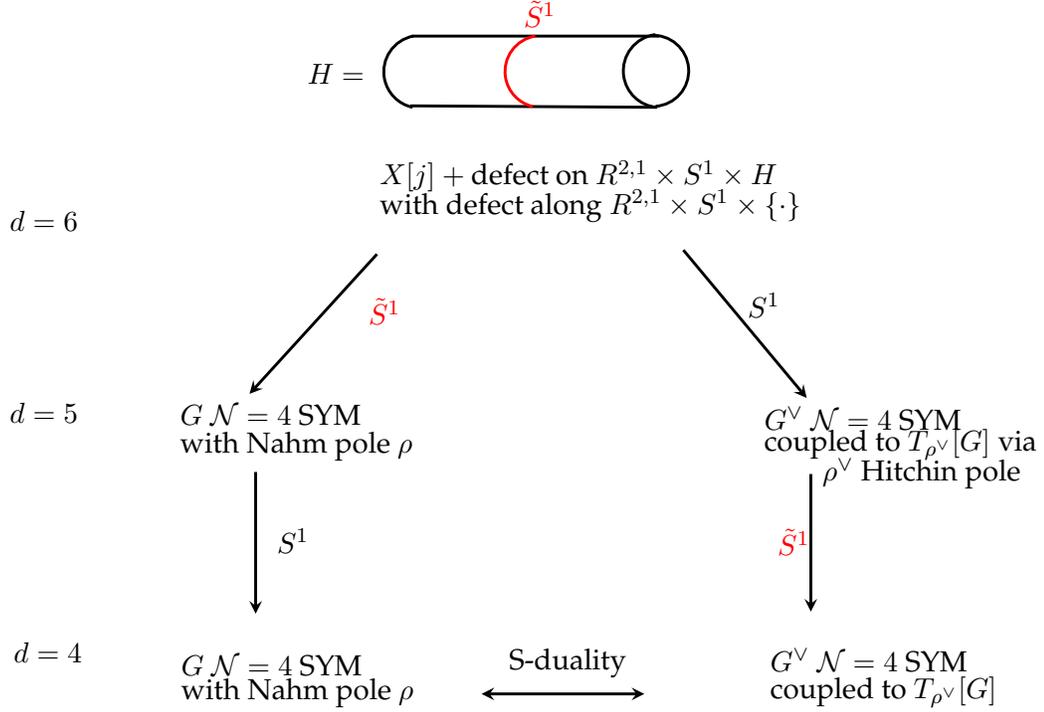

The existence of the regular codimension two defects of the six dimensional $\mathscr{X}[j]$ theory can be related to the existence of certain 1/2 BPS boundary conditions of $\mathcal{N}=4$ SYM in four dimensions. This can be inferred by considering the following setup (see Fig \ref{nahmhitchin}) . In six dimensions, formulate theory $\mathscr{X}[j]$ on a spacetime of the form $\mathbb{R}^{1,2} \times \mathbb{S}^1 \times H$, where $H$ is a two dimensional disc with a cigar metric. One can reduce to four dimensions in two different ways. First, we reduce along the circle fiber $\tilde{S}^1$ of the cigar $H$ to get five dimensional $\mathcal{N}=4$ SYM with gauge group $G$ (a compact group) together with a boundary condition that is specified by a Nahm pole. If a twist by Dynkin diagram automorphism of $j$ is included in the $\tilde{S}^1$ reduction, then $G$ is the compact group corresponding to the `folded' Dynkin diagram. This allows for the appearance of non-simply laced $G$. If the twist is not included, $G$ is the compact group corresponding to complex lie algebra $j$ and is thus simply laced. Reducing further on the circle $S^1$, one gets four dimensional $\mathcal{N}=4$ SYM with gauge group $G$ formulated on a four manifold $M_4$ that is topologically $\mathbb{R}^{1,2}\times \mathbb{R}^+$ together with a boundary condition that is labeled by the same Nahm pole $\rho$. If the two dimensional reductions are done in the opposite order, one gets $G^\vee$ $\mathcal{N}=4$ theory with a boundary condition that involves coupling to a boundary three dimensional $\mathcal{N}=4$ theory called $T_{\rho^\vee}[G]$. Let us pick a co-ordinate $y$ for $\mathbb{R}^+$ direction such that $y=0$ forms the boundary of $M_4$. The presence of a Nahm pole boundary condition of type $\rho$ for the four dimensional $\mathcal{N}=4$ theory with gauge group $G$ implies (by definition), the following equations for three out of the six scalars in the theory,
\begin{eqnarray}
\frac{dX^i}{dy} &=& \epsilon^{ijk} [X^j,X^k]  \\ 
X^i &=& \frac{\rho(\tau^i)}{y} \nonumber,
\label{Nahmequations}
\end{eqnarray}
where $\rho$ denotes an embedding of $\mathfrak{sl}_2$ into the complex lie algebra $\mathfrak{g}$ and $\tau^i$ are the standard generators of $\mathfrak{sl}_2$. They obey $[\tau^1,\tau^3]=2\tau^2,[\tau^2,\tau^1]=\tau^1,[\tau^2,\tau^3]=-\tau^3$. Here, the three scalars have been arranged into a vector $X$ and they transform in the three dimensional representation of the $SO(3)_X$ part of the $R$ symmetry group. There is a corresponding set of equations that are obeyed by the gauge fields in the theory. When the $\mathfrak{sl}_2$ embedding is trivial, the gauge field obey the usual Dirichlet boundary conditions. Hence, the most general Nahm pole boundary conditions can be viewed as generalized Dirichlet boundary conditions.

In \cite{Gaiotto:2008sa}, Gaiotto-Witten constructed a vast class of 1/2 BPS conformal boundary conditions in four dimensional $\mathcal{N}=4$ SYM theory formulated on $\mathbb{R}^{1,2}\times \mathbb{R}^+$ of which the Nahm pole boundary conditions form a small but very interesting subset.  Their classification was by associating a triple $(\mathcal{O}^\rho, H,\mathcal{B})$ to every boundary condition, where $\mathcal{O}^\rho$ is a nilpotent orbit in the lie algebra $\mathfrak{g}$ (by the Jacobson-Morozov theorem, this is equivalent to a choice of an embedding $\rho: \mathfrak{sl}_2 \rightarrow \mathfrak{g}$) and $H$ is a subgroup of the centralizer $Z_{\mathfrak{g}}(\rho)$ of the associated $sl_2$ triple and $\mathcal{B}$ is a three dimensional SCFT with eight supercharges. 

The translation from the indexing set of triples to the boundary condition is as follows. First, one imposes Nahm pole boundary conditions for a triplet of the scalars (labeled $X^i$) as in Eqns \ref{Nahmequations}. The second datum implies that one considers a decomposition of the lie algebra into $\mathfrak{h} \oplus \mathfrak{h}^\perp$ and Higgs the theory down to $H$ symmetry at the boundary. The third datum in the triple involves choosing a a three dimensional $\mathcal{N}=4$ SCFT $\mathcal{B}$ with $H$ global symmetry and gauging the global symmetry and coupling the theory at the boundary to the four dimensional theory.

In \cite{Gaiotto:2008ak}, they further studied the action of S-duality on this vast class of boundary conditions. One of the important instances of this duality of boundary conditions is the case of a pure Nahm pole boundary condition of type $\rho$. The S-dual of this boundary condition happens to be a boundary condition in the $G^\vee$ theory that corresponds to coupling to the boundary theory $T_{\rho^\vee}$. In terms of the indexing set of triple, this can be stated as,
\begin{equation}
(\mathcal{O}^\rho,Id,\varnothing )  \leftrightarrow_{S-dual} (\mathcal{O}^0,  G^\vee, T_{\rho^\vee}[G]) .
\label{dualitymap}
\end{equation}
In order the make precise the idea of `coupling at the boundary', one needs to recall some facts about $\mathcal{N}=4$ SCFTs in three dimensions. In general, these theories admit atleast two quantum moduli spaces of vacua called the Higgs branch and the Coulomb branch. The $\mathcal{N}=4$ supersymmetry of the theory ensures that both of these are hyperkahler spaces. In theories with Lagrangian descriptions, the Higgs can be described as a hyper-Kahler quotient and the metric is not `quantum corrected'. The Coulomb branch, on the other hand \textit{is} changed by quantum corrections and is, generically, much harder to understand. For theories without Lagrangian descriptions, one is, \textit{a priori} at a loss to describe to either of the branches. The $\mathcal{N}=4$ theories in four dimensional admit a duality called \textit{three dimensional mirror symmetry} that, among other things, exchanges the Higgs and Coulomb branches. For the theories denoted by $T_{\rho^\vee}[G]$, the Coulomb branches are specific strata inside the nilpotent cone of $\mathfrak{g}$ ($\mathcal{N}_{\mathfrak{g}}$) called Slodowy slices and the Higgs branches are certain strata inside the nilpotent cone of $\mathfrak{g}^\vee$ ($\mathcal{N}_{\mathfrak{g}^\vee}$) corresponding to nilpotent orbit closures. So, the coupling at the boundary for the $G^\vee$ theory is to be understood to proceed via the gauging of the global symmetry on the Higgs branch of the $T_{\rho^\vee}[G]$ theory. This is an effective IR description of what it means to `couple to a boundary theory'. On the $G$ side, the space of solutions to the Nahm equations can be non-trivial and if this is the case, the four dimensional theory is be interpreted to have a moduli space of vacua given by this space of solutions. Not coincidentally, these solutions are exactly the Slodowy slices that arise as Coulomb branches of the $T_{\rho^\vee}[G]$ theories. The fact that the same moduli spaces have different realizations on the $G$ and $G^\vee$ is the justification for the S-duality map in \ref{dualitymap}. 

Now, note that on the $G$ side of the duality, the classical gauge fields are $G$ valued. So, if one were to start with the data describing the boundary condition on the $G$ side, it seems reasonable that one is able to directly conclude something about part of the vacuum moduli space of $T_{\rho^\vee}[G]$ that is a stratum inside $\mathcal{N}_{\mathfrak{g}}$. This, however, constitutes `classical' data for the mirror dual of $T_{\rho^\vee}[G]$ and not for $T_{\rho^\vee}[G]$ itself. This mirror dual is usually denoted by $T^{\rho}[G]$.  For this reason, S-duality of boundary conditions in the four dimensional $\mathcal{N}=4$ theory is inextricably connected with mirror symmetry for three dimensional $\mathcal{N}=4$ SCFTs. 

In order to identify precisely which $T_{\rho^\vee}[G]$ appear as duals of a given Nahm pole boundary conditions, Gaiotto-Witten proposed using the dimension of the moduli space as an invariant. It can be shown that for $G=SU(N)$, this is an adequate invariant and matching just the dimensions of moduli spaces on both sides leads to a complete specification of the duality map. However, for arbitrary $G$, this is inadequate. This motivates the search for a finer invariant of the moduli spaces. In the following section, one such invariant is discussed.

\section{Springer correspondence and the Springer invariant}
\label{springer}
The classical Springer correspondence is a construction that relates the geometric world of nilpotent orbits and Slodowy slices to the algebraic world of irreducible representations of the Weyl group. What follows is an extremely short review of a rich subject. The reader is referred to \cite{Balasubramanian:2014jca} for a longer discussion. The starting point in the construction is the existence of the \textit{Springer resolution} $\mu : \tilde{\mathcal{N}} \rightarrow \mathcal{N}$. This is a simultaneous resolution of the singularities of $\mathcal{N}$. From a physical perspective, the \textit{existence} of such a resolution of the vacuum moduli spaces is related to the existence of Fayet-Iliopoulos and Mass parameters in the $\mathcal{N}=4$ theory. Typically, turning on real values for either the F.I or mass parameters leads to a resolution of the Higgs branch of a given theory $T_{\rho^\vee}[G]$ or Higgs branch of its mirror $T^\rho[G]$. 

The existence of such a map $\mu$ ties together topological properties of the resolved space $\mathcal{N}$ and the singular space $\mathcal{N}$ in fairly intricate ways. An example of such a relationship is the decomposition theorem. The Springer correspondence can be viewed as a small but powerful part of the results obtained by applying the decomposition theorem to the particular setting of the Springer resolution (for a review, see \cite{de2009decomposition}). To be more specific, let $e$ be a representative of a nilpotent orbit $\mathcal{O}$. the Springer correspondence is a statement about how the top dimensional cohomology of the Springer fiber $\mathcal{B}_e (=\mu^{-1}(e))$ decomposes as a module for the Weyl group. The correspondence can be encapsulated in an injective map,
\begin{equation}
 Sp  : Irr(W) \rightarrow \tilde{\mathcal{O}},
 \label{correspondence}
\end{equation}
 
where $Irr(W)$ is the set of all irreducible representation of the Weyl group associated to $\mathfrak{g}$ and $\tilde{\mathcal{O}}$ is the set of all pairs $(\mathcal{O},\chi)$, where $\mathcal{O}$ is a nilpotent orbit and $\chi$ is an irreducible representation of the component group $A(\mathcal{O})$ associated to the nilpotent orbit $\mathcal{O}$. The component group is defined to be the quotient $C_G(e)/C_G^0(e)$, where $C_G(e)$ is the centralizer of the nilpotent element in the group and $C_G^0(e)$ is its connected component. It is often useful to consider the inverse of this map and denote it as $Sp^{-1} :\tilde{\mathcal{O}}\rightarrow Irr(W)$. Some care is required in using the inverse since the map $Sp$ is guaranteed only to be injective. In what follows, we will be able to use the inverse without worrying about this subtlety. 
 Now, since both nilpotent orbits and Slodowy slices are resolved under the Springer map, the constructions like the Springer correspondence apply to both kinds of strata. Physically, this means one can separately attach a Higgs branch Springer invariant (denoted by $r$) and a Coulomb branch Springer invariant (denoted by $\bar{r}$) for theories like $T_{\rho^\vee}[G]$. Outside of case $G=SU(N)$, there is an asymmetry between the properties of the two possible Springer invariants. This asymmetry singles out one of them as being the more useful Springer invariant. In this note, the Coulomb branch Springer invariant of $T_{\rho^\vee}[G]$ is the one that plays this role. \footnote{In the work \cite{Balasubramanian:2014jca}, all statements were made using the $T^\rho[G]$ theories and hence the same invariant was called the Higgs branch Springer invariant.}

\section{Classification via Symmetry breaking}
\label{Classification}

In this Section, we will see that the invariant constructed in Sec \ref{springer} is strictly smarter than the dimension of the associated vacuum moduli space. In particular, it allows for an elegant description of which exact strata pair up to form the Higgs and Coulomb branches of $T_{\rho^\vee}[G]$. Further, it allows a re-organization of the classification of the associated boundary conditions in the four dimensional theory. The seemingly ad-hoc nature (from a physical standpoint) of the association of the invariant is compensated by the incredibly simple and powerful statements that one can make using this invariant. 

As discussed in the earlier section, coupling to $T_{\rho^\vee}[G]$ in the boundary is via a nilpotent representative $\rho^\vee$ of the nilpotent orbit that is identified as the Higgs branch of the theory. Now, let us think of the coupling to the boundary theory under possible symmetry breaking in the bulk. Note that this is a \textit{different} situation from the one where we consider symmetry breaking only at the boundary. The symmetric breaking is triggered by a vev $m$ for the bulk scalar and unbroken gauge group is the centralizer $Z_{\mathfrak{g}^\vee}(m)$ of the semi-simple element $m \in \mathfrak{g}^\vee$.  Let the connected part of the unbroken gauge group be $L^\vee$. Now, a necessary condition for a coupling via $\rho^\vee$ to make sense is that there should be enough unbroken gauge symmetry at the boundary so that $\rho^\vee$ is still a nilpotent representative of some nilpotent orbit in the lie algebra $l^\vee$. Clearly, for any non-zero nilpotent element $\rho^\vee$, making $l^\vee$ too small would violate this requirement. The most that one can do is demand that $\rho^\vee$ correspond to a `a very big orbit' in $l^\vee$. Loosely speaking, it is so big that it can't fit into a smaller subalgebra. A mathematically precise way of enforcing this notion is to demand that $\rho^\vee$ is a distinguished orbit in $l^\vee$. 

Those familiar with the structure theory of nilpotent orbits will immediately recognize the close similarity between what has just been said and Bala-Carter classification of nilpotent orbits. There is, however, one extremely important difference. In the work of Bala-Carter, the problem of classifying nilpotent orbits is mapped to the problem of classifying distinguished nilpotent orbits in \textit{Levi} subalgebras\footnote{Equivalently, one looks at distinguished nilpotent orbits in parabolic subalgebras. Each parabolic subalgebra has a canonical Levi decomposition whose `Levi part' is called a Levi subalgebra.}. Note that in our discussion of symmetry breaking, no mention of $l^\vee$ being related to Levi subalgebra was mentioned. The relationship that one can hope for is that the lie algebra of the connected component of the unbroken gauge group ($l^\vee$, in current notation) occurs as the semi-simple part of a Levi subalgebra. Such a relationship indeed exists for groups of type A. But, outside of groups of type A, the above statement is $\textit{not true}$ in general. For groups outside of type A, only a proper subset of the centralizers of semi-simple elements are related to Levi subalgebras. This is part of what makes this extremely interesting. A study of distinguished nilpotent orbits in these more general centralizers was done by Sommers in \cite{sommers1998generalization} and this extends the work of Bala-Carter. It is this extended form of the Bala-Carter classification, one that can be termed Bala-Carter-Sommers classification, that is relevant for the above discussion treating the classification of defects in conjunction with symmetry breaking patterns. For a given theory $T_{\rho^\vee}[G]$, let $e^\vee$ be the nilpotent representative of the orbit that represents the Higgs branch and let $(\mathfrak{l}^\vee,e^\vee)$ be an associated Sommers pair. 

Now, let us consider a \textit{distinguished} symmetry breaking in the bulk of the $G^\vee$ theory. This means the following. One picks a vev $m$ for the scalar in the vector multiplet such that the RG flow triggered by the vev $m$ lands in the theory with a unbroken gauge symmetry whose connected component is the group $L^\vee$. In other words, the RG flow is picked so that in the UV, one starts with a random nilpotent orbit in $g$ and in the IR, one ends up recovering the Sommers pair $(l^\vee,e^\vee)$. The fact that the Sommers pair is a characterization of the Higgs branch of $T_{\rho^\vee}[G]$ is obvious from the way it has been constructed. It will, in fact, turn out to identify it uniquely after we impose a subtle equivalence relation on Sommers pairs (this is discussed below). But, a natural question is how does one relate the Coulomb branch of $T_{\rho^\vee}[G]$ to their corresponding Sommers pair of the Higgs branch. It turns out that the following matching conditions for the Springer invariants specifies this relationship completely,
\begin{eqnarray}
s &=& \epsilon_{\mathfrak{l}} \times Sp^{-1} [\mathfrak{l}^\vee]  \\
Sp^{-1}[\mathfrak{g},\mathcal{O}_N] &=& j_{W[\mathfrak{l}^\vee]}^{W[\mathfrak{g}^\vee]} (s)  \nonumber,
\label{matching}
\end{eqnarray}

where $\mathcal{O}_N$ is the nilpotent orbit that enters the description of the Nahm boundary condition (and is called the Nahm datum in Fig \ref{nahmhitchin}) and $\mathcal{O}_H$ is the nilpotent orbit that corresponds to the Higgs branch of $T_{\rho^\vee}[G]$ and is called the Hitchin datum in Fig \ref{nahmhitchin}. The invariant that was previously dubbed the Coulomb branch Springer invariant  denoted by $\bar{r}$ is the same as  $Sp^{-1}[\mathfrak{g},\mathcal{O}_N]$ in the matching condition above.  For the class of boundary conditions being considered, it is sufficient to restrict to the case where $\mathcal{O}_H$ is a special nilpotent orbit. The operation denoted by $j$ is a functor between the irreducible representations of the Weyl group $W[\mathfrak{l}^\vee]$ and that of the Weyl group $W[\mathfrak{g^\vee}]$. It is a truncated version of the usual induction functor and the nature of its action is explained in greater detail in \cite{Balasubramanian:2014jca}.

Now, for a given nilpotent representative $e^\vee$, there could be different choices (non conjugate) for $\mathfrak{l}^\vee$ such that $(\mathfrak{l}^\vee,e^\vee)$ is a Sommers pair. What does this indicate ? This indicates that one is dealing with two different theories $T^\rho[G]$ and $T^{\rho'}[G]$ whose associated Coulomb branches correspond to the Sommers pairs $(\mathfrak{l}^\vee,e^\vee)$ and $(\mathfrak{l}^{\vee'},e^\vee)$. The number of such distinct Sommers pairs associated to a given special nilpotent orbit is indicative of the size of its associated $\overline{A}(O)$ group.  

One of the esoteric features of this setup is the appearance of the group $\overline{A}(O)$.\footnote{To paint a more complete picture, one actually needs to involve more data about conjugacy classes/subgroups in $\overline{A}(O)$. For this, we refer to \cite{Balasubramanian:2014jca,Chacaltana:2012zy}} In several cases, it is the same as the group $A(O)$ that we encountered before, but it is in general a smaller group. This group is a quotient of the component group $A(O)$. It was originally encountered by Lusztig in his work \cite{lusztig1984characters} and is thus sometimes called ``Lusztig's quotient''. In terms of the matching conditions for the Springer invariant, the appearance of the smaller group can be understood in the following way. For some $G^\vee$, there exist different $L^\vee$ (not conjugate to each other) for which the output of $j$-induction (invoked as in \ref{matching}) happens to be the same irrep of $W[\mathfrak{g^\vee}]\cong W[\mathfrak{g}]$. Now, define an equivalence relation on Sommers pairs for a particular nilpotent orbit such that they are identified if the output of $j$-induction is the same. The Sommers pairs, together with equivalence relation now give a characterization of the group $\overline{A}(O)$. As an example of the cases where $\overline{A}(O) \neq A(O)$ , we list here two such instances for $G^\vee=E_8$.
\begin{center}
\begin{tabular}{|c|c|c|}
\hline 
$\mathcal{O}_H$ & $A(O)$ & $\overline{A}(O)$ \\ \hline 
$E_7(a_4)$ & $S_2$ & $1$ \\ 
$E_8(b_6)$ & $S_3$ & $S_2$ \\ 
\hline 
\end{tabular} 
\end{center}

\subsection{A pair of functors}
In the above discussion, we have really encountered two functors between classes of boundary conditions. The first is distinguished symmetry breaking in the presence of a boundary condition. Let us called this the functor $s$. The functor $s$ goes from theory with gauge symmetry $G^\vee$ to the theory with gauge symmetry $L^\vee$. The second is the functor that goes between boundary conditions in theory with gauge symmetry $L^\vee$ and the theory with gauge symmetry $G^\vee$. Let us call this functor $i$. These functors can be viewed as being analogous to the restrict and induce functors (which form an adjoint pair) between the categories of representation of a finite group $\mathcal{F}$ and a subgroup $\mathcal{H} \subset \mathcal{F}$. In terms of the associated Springer invariants, the functors encountered above are directly the $Res$ and $j$ induction functors in the study of Weyl group representations. Note however that one is dealing here with the class of boundary conditions in a four dimensional quantum field theory. This is morally modeled by an associated 3-category. So, a complete understanding of the pair of functors $(s,i)$ should be developed in this setting.

\subsection{Mirror symmetry and Symplectic duality}
We have seen that S-duality of boundary conditions in four dimensional $\mathcal{N}=4$ has intimate connections with mirror symmetry for certain associated three dimensional SCFTS. From a mathematical perspective, it turns out that this 3d mirror symmetry has close connections with the idea of symplectic duality in the sense of BLPW \cite{braden2014quantizations}. According to BLPW, two conical symplectic resolutions $\mu:M \rightarrow M_0 $ and $\mu' : M' \rightarrow M'_0$ are defined to be symplectic duals of each other if certain associated categories are related to each other via Koszul duality. From a physical perspective, the two resolved spaced $M$ and $M'$ are resolved Higgs branches of two different 3d SCFTs which are mirror duals of each other. The case of $T^\rho[G]$ and $T_{\rho^\vee}[G]$ is a particular instance of such a mirror pair and the relevant Higgs branches are, respectively, Slodowy slices and nilpotent orbit closures. The relevant symplectic resolution is the Springer resolution discussed in Section \ref{springer}. This relationship has several consequences, one of which is a relationship between pieces of the cohomology of the resolved spaces $M$ and $M'$ (discussed in \cite{braden2014quantizations,proudfoot2014poisson}). It would be interesting to understand the precise relationship between the cohomological consequences of symplectic duality and the matching condition that occurs in Sec \ref{Classification}. 

The relationship at the level of categories between the mathematical definition of symplectic duality and 3d mirror symmetry  has been studied recently for many 3d $\mathcal{N}=4$ theories \cite{dimoftetalk}. The braid group actions that play an important role in the duality between categories yield Weyl group actions at the level of cohomology upon suitable `decategorification'.

\vskip 1cm

\bibliographystyle{JHEP.bst}
\bibliography{springer.bbl}
\end{document}

%% file: figs/nahmhitchin.tex
\ifx\du\undefined
  \newlength{\du}
\fi
\setlength{\du}{11\unitlength}
\begin{tikzpicture}
\pgftransformxscale{1.000000}
\pgftransformyscale{-1.000000}
\definecolor{dialinecolor}{rgb}{0.000000, 0.000000, 0.000000}
\pgfsetstrokecolor{dialinecolor}
\definecolor{dialinecolor}{rgb}{1.000000, 1.000000, 1.000000}
\pgfsetfillcolor{dialinecolor}
\pgfsetlinewidth{0.100000\du}
\pgfsetdash{}{0pt}
\pgfsetdash{}{0pt}
\pgfsetbuttcap
{
\definecolor{dialinecolor}{rgb}{0.000000, 0.000000, 0.000000}
\pgfsetfillcolor{dialinecolor}
\definecolor{dialinecolor}{rgb}{0.000000, 0.000000, 0.000000}
\pgfsetstrokecolor{dialinecolor}
\pgfpathmoveto{\pgfpoint{24.241458\du}{-0.637613\du}}
\pgfpatharc{259}{102}{1.250313\du and 1.250313\du}
\pgfusepath{stroke}
}
\pgfsetlinewidth{0.100000\du}
\pgfsetdash{}{0pt}
\pgfsetdash{}{0pt}
\pgfsetbuttcap
{
\definecolor{dialinecolor}{rgb}{0.000000, 0.000000, 0.000000}
\pgfsetfillcolor{dialinecolor}
\definecolor{dialinecolor}{rgb}{0.000000, 0.000000, 0.000000}
\pgfsetstrokecolor{dialinecolor}
\draw (24.191421\du,-0.637605\du)--(32.691421\du,-0.637605\du);
}
\pgfsetlinewidth{0.100000\du}
\pgfsetdash{}{0pt}
\pgfsetdash{}{0pt}
\pgfsetbuttcap
{
\definecolor{dialinecolor}{rgb}{0.000000, 0.000000, 0.000000}
\pgfsetfillcolor{dialinecolor}
\definecolor{dialinecolor}{rgb}{0.000000, 0.000000, 0.000000}
\pgfsetstrokecolor{dialinecolor}
\draw (24.141421\du,1.762395\du)--(32.591421\du,1.812395\du);
}
\pgfsetlinewidth{0.100000\du}
\pgfsetdash{}{0pt}
\pgfsetdash{}{0pt}
\pgfsetbuttcap
{
\definecolor{dialinecolor}{rgb}{0.000000, 0.000000, 0.000000}
\pgfsetfillcolor{dialinecolor}
\definecolor{dialinecolor}{rgb}{0.000000, 0.000000, 0.000000}
\pgfsetstrokecolor{dialinecolor}
\pgfpathmoveto{\pgfpoint{32.570746\du}{-0.687610\du}}
\pgfpatharc{263}{94}{1.257104\du and 1.257104\du}
\pgfusepath{stroke}
}
\pgfsetlinewidth{0.100000\du}
\pgfsetdash{}{0pt}
\pgfsetdash{}{0pt}
\pgfsetbuttcap
{
\definecolor{dialinecolor}{rgb}{0.000000, 0.000000, 0.000000}
\pgfsetfillcolor{dialinecolor}
\definecolor{dialinecolor}{rgb}{0.000000, 0.000000, 0.000000}
\pgfsetstrokecolor{dialinecolor}
\pgfpathmoveto{\pgfpoint{32.575817\du}{1.777730\du}}
\pgfpatharc{86}{-94}{1.222961\du and 1.222961\du}
\pgfusepath{stroke}
}
\pgfsetlinewidth{0.100000\du}
\pgfsetdash{}{0pt}
\pgfsetdash{}{0pt}
\pgfsetbuttcap
{
\definecolor{dialinecolor}{rgb}{1.000000, 0.000000, 0.000000}
\pgfsetfillcolor{dialinecolor}
\definecolor{dialinecolor}{rgb}{1.000000, 0.000000, 0.000000}
\pgfsetstrokecolor{dialinecolor}
\pgfpathmoveto{\pgfpoint{28.441457\du}{-0.637611\du}}
\pgfpatharc{261}{103}{1.235781\du and 1.235781\du}
\pgfusepath{stroke}
}
\pgfsetlinewidth{0.100000\du}
\pgfsetdash{}{0pt}
\pgfsetdash{}{0pt}
\pgfsetbuttcap
{
\definecolor{dialinecolor}{rgb}{0.000000, 0.000000, 0.000000}
\pgfsetfillcolor{dialinecolor}
\pgfsetarrowsend{stealth}
\definecolor{dialinecolor}{rgb}{0.000000, 0.000000, 0.000000}
\pgfsetstrokecolor{dialinecolor}
\draw (22.994618\du,6.866807\du)--(18.610556\du,11.675133\du);
}
\pgfsetlinewidth{0.100000\du}
\pgfsetdash{}{0pt}
\pgfsetdash{}{0pt}
\pgfsetbuttcap
{
\definecolor{dialinecolor}{rgb}{0.000000, 0.000000, 0.000000}
\pgfsetfillcolor{dialinecolor}
\pgfsetarrowsend{stealth}
\definecolor{dialinecolor}{rgb}{0.000000, 0.000000, 0.000000}
\pgfsetstrokecolor{dialinecolor}
\draw (33.530509\du,6.725386\du)--(37.773149\du,11.816554\du);
}
\definecolor{dialinecolor}{rgb}{0.000000, 0.000000, 0.000000}
\pgfsetstrokecolor{dialinecolor}
\node[anchor=west] at (9.983853\du,5.735436\du){$d=6$};
\definecolor{dialinecolor}{rgb}{0.000000, 0.000000, 0.000000}
\pgfsetstrokecolor{dialinecolor}
\node[anchor=west] at (9.983853\du,12.311529\du){$d=5$};
\definecolor{dialinecolor}{rgb}{0.000000, 0.000000, 0.000000}
\pgfsetstrokecolor{dialinecolor}
\node[anchor=west] at (10.125274\du,20.584679\du){$d=4$};
\pgfsetlinewidth{0.100000\du}
\pgfsetdash{}{0pt}
\pgfsetdash{}{0pt}
\pgfsetbuttcap
{
\definecolor{dialinecolor}{rgb}{0.000000, 0.000000, 0.000000}
\pgfsetfillcolor{dialinecolor}
\pgfsetarrowsend{stealth}
\definecolor{dialinecolor}{rgb}{0.000000, 0.000000, 0.000000}
\pgfsetstrokecolor{dialinecolor}
\draw (18.822688\du,14.362139\du)--(18.822688\du,19.241176\du);
}
\pgfsetlinewidth{0.100000\du}
\pgfsetdash{}{0pt}
\pgfsetdash{}{0pt}
\pgfsetbuttcap
{
\definecolor{dialinecolor}{rgb}{0.000000, 0.000000, 0.000000}
\pgfsetfillcolor{dialinecolor}
\pgfsetarrowsend{stealth}
\definecolor{dialinecolor}{rgb}{0.000000, 0.000000, 0.000000}
\pgfsetstrokecolor{dialinecolor}
\draw (37.914571\du,14.432850\du)--(37.914571\du,19.170465\du);
}
\definecolor{dialinecolor}{rgb}{0.000000, 0.000000, 0.000000}
\pgfsetstrokecolor{dialinecolor}
\node[anchor=west] at (37.773149\du,18.958333\du){};
\definecolor{dialinecolor}{rgb}{0.000000, 0.000000, 0.000000}
\pgfsetstrokecolor{dialinecolor}
\node[anchor=west] at (27.944365\du,-1.759896\du){};
\definecolor{dialinecolor}{rgb}{0.000000, 0.000000, 0.000000}
\pgfsetstrokecolor{dialinecolor}
\node[anchor=west] at (27.661522\du,-1.406342\du){$\txtred{\tilde{S}^1}$};
\definecolor{dialinecolor}{rgb}{0.000000, 0.000000, 0.000000}
\pgfsetstrokecolor{dialinecolor}
\node[anchor=west] at (22.711775\du,4.179801\du){$X[j] + \text{defect on } R^{2,1} \times S^1 \times H$};
\definecolor{dialinecolor}{rgb}{0.000000, 0.000000, 0.000000}
\pgfsetstrokecolor{dialinecolor}
\node[anchor=west] at (22.711775\du,5.200801\du){with defect along $R^{2,1}\times S^1 \times \{ \cdot \}$};
\definecolor{dialinecolor}{rgb}{0.000000, 0.000000, 0.000000}
\pgfsetstrokecolor{dialinecolor}
\node[anchor=west] at (15.852839\du,12.382240\du){$G$ $\mathcal{N}=4$ SYM};
\definecolor{dialinecolor}{rgb}{0.000000, 0.000000, 0.000000}
\pgfsetstrokecolor{dialinecolor}
\node[anchor=west] at (15.852839\du,13.482240\du){with Nahm pole $\rho$};
\definecolor{dialinecolor}{rgb}{0.000000, 0.000000, 0.000000}
\pgfsetstrokecolor{dialinecolor}
\node[anchor=west] at (35.934672\du,12.523661\du){$G^\vee$ $\mathcal{N}=4$ SYM};
\definecolor{dialinecolor}{rgb}{0.000000, 0.000000, 0.000000}
\pgfsetstrokecolor{dialinecolor}
\node[anchor=west] at (35.934672\du,13.423661\du){coupled to $T_{\rho^\vee}[G]$ via};
\node[anchor=west] at (37.934672\du,14.423661\du){$\rho^\vee$ Hitchin pole};
\pgfsetlinewidth{0.100000\du}
\pgfsetdash{}{0pt}
\pgfsetdash{}{0pt}
\pgfsetbuttcap
{
\definecolor{dialinecolor}{rgb}{0.000000, 0.000000, 0.000000}
\pgfsetfillcolor{dialinecolor}
\pgfsetarrowsstart{stealth}
\pgfsetarrowsend{stealth}
\definecolor{dialinecolor}{rgb}{0.000000, 0.000000, 0.000000}
\pgfsetstrokecolor{dialinecolor}
\draw (26.590381\du,22.0\du)--(32.190381\du,22.0\du);
}
\definecolor{dialinecolor}{rgb}{0.000000, 0.000000, 0.000000}
\pgfsetstrokecolor{dialinecolor}
\node[anchor=west] at (15.888194\du,21.014453\du){$G$ $\mathcal{N}=4$ SYM};
\definecolor{dialinecolor}{rgb}{0.000000, 0.000000, 0.000000}
\pgfsetstrokecolor{dialinecolor}
\node[anchor=west] at (15.888194\du,21.974453\du){with Nahm pole $\rho$};
\definecolor{dialinecolor}{rgb}{0.000000, 0.000000, 0.000000}
\pgfsetstrokecolor{dialinecolor}
\node[anchor=west] at (36.146804\du,20.837676\du){$G^\vee$ $\mathcal{N}=4$ SYM};
\definecolor{dialinecolor}{rgb}{0.000000, 0.000000, 0.000000}
\pgfsetstrokecolor{dialinecolor}
\node[anchor=west] at (36.146804\du,21.937676\du){coupled to $T_{\rho^\vee}[G]$};
\definecolor{dialinecolor}{rgb}{0.000000, 0.000000, 0.000000}
\pgfsetstrokecolor{dialinecolor}
\node[anchor=west] at (27.166548\du,20.943742\du){S-duality};
\definecolor{dialinecolor}{rgb}{0.000000, 0.000000, 0.000000}
\pgfsetstrokecolor{dialinecolor}
\node[anchor=west] at (22.322866\du,8.836389\du){$\txtred{\tilde{S}^1}$};
\definecolor{dialinecolor}{rgb}{0.000000, 0.000000, 0.000000}
\pgfsetstrokecolor{dialinecolor}
\node[anchor=west] at (28.156497\du,-1.647789\du){};
\definecolor{dialinecolor}{rgb}{0.000000, 0.000000, 0.000000}
\pgfsetstrokecolor{dialinecolor}
\node[anchor=west] at (27.696878\du,-1.577078\du){};
\definecolor{dialinecolor}{rgb}{0.000000, 0.000000, 0.000000}
\pgfsetstrokecolor{dialinecolor}
\node[anchor=west] at (27.979720\du,-1.577078\du){};
\definecolor{dialinecolor}{rgb}{0.000000, 0.000000, 0.000000}
\pgfsetstrokecolor{dialinecolor}
\node[anchor=west] at (29.005025\du,-1.683144\du){};
\definecolor{dialinecolor}{rgb}{0.000000, 0.000000, 0.000000}
\pgfsetstrokecolor{dialinecolor}
\node[anchor=west] at (35.368986\du,8.627503\du){$S^1$};
\definecolor{dialinecolor}{rgb}{0.000000, 0.000000, 0.000000}
\pgfsetstrokecolor{dialinecolor}
\node[anchor=west] at (19.176241\du,16.734482\du){$S^{1}$};
\definecolor{dialinecolor}{rgb}{0.000000, 0.000000, 0.000000}
\pgfsetstrokecolor{dialinecolor}
\node[anchor=west] at (36.394291\du,16.776909\du){$\txtred{\tilde{S}^{1}}$};
\definecolor{dialinecolor}{rgb}{0.000000, 0.000000, 0.000000}
\pgfsetstrokecolor{dialinecolor}
\node[anchor=west] at (20.226851\du,0.643790\du){$H =$ };
\definecolor{dialinecolor}{rgb}{0.000000, 0.000000, 0.000000}
\pgfsetstrokecolor{dialinecolor}
\node[anchor=west] at (29.818198\du,3.961126\du){};
\end{tikzpicture}

%% file: springer.bbl
\providecommand{\href}[2]{#2}\begingroup\raggedright\begin{thebibliography}{10}

\bibitem{Gaiotto:2009we}
D.~Gaiotto, {\it {N=2 dualities}},  {\em JHEP} {\bf 1208} (2012) 034,
  [\href{http://xxx.lanl.gov/abs/0904.2715}{{\tt arXiv:0904.2715}}].

\bibitem{Gaiotto:2009hg}
D.~Gaiotto, G.~W. Moore, and A.~Neitzke, {\it {Wall-crossing, Hitchin Systems,
  and the WKB Approximation}},  \href{http://xxx.lanl.gov/abs/0907.3987}{{\tt
  arXiv:0907.3987}}.

\bibitem{Balasubramanian:2014jca}
A.~Balasubramanian, {\it {Describing codimension two defects}},  {\em JHEP}
  {\bf 1407} (2014) 095, [\href{http://xxx.lanl.gov/abs/1404.3737}{{\tt
  arXiv:1404.3737}}].

\bibitem{Gaiotto:2008sa}
D.~Gaiotto and E.~Witten, {\it {Supersymmetric Boundary Conditions in N=4 Super
  Yang-Mills Theory}},  {\em J.Statist.Phys.} {\bf 135} (2009) 789--855,
  [\href{http://xxx.lanl.gov/abs/0804.2902}{{\tt arXiv:0804.2902}}].

\bibitem{Gaiotto:2008ak}
D.~Gaiotto and E.~Witten, {\it {S-Duality of Boundary Conditions In N=4 Super
  Yang-Mills Theory}},  {\em Adv.Theor.Math.Phys.} {\bf 13} (2009) 721,
  [\href{http://xxx.lanl.gov/abs/0807.3720}{{\tt arXiv:0807.3720}}].

\bibitem{de2009decomposition}
M.~de~Cataldo and L.~Migliorini, {\it The decomposition theorem, perverse
  sheaves and the topology of algebraic maps},  {\em Bulletin of the American
  Mathematical Society} {\bf 46} (2009), no.~4 535--633.

\bibitem{sommers1998generalization}
E.~Sommers, {\it A generalization of the bala-carter theorem for nilpotent
  orbits},  {\em International Mathematics Research Notices} {\bf 1998} (1998),
  no.~11 539--562.

\bibitem{Chacaltana:2012zy}
O.~Chacaltana, J.~Distler, and Y.~Tachikawa, {\it {Nilpotent orbits and
  codimension-two defects of 6d N=(2,0) theories}},
  \href{http://xxx.lanl.gov/abs/1203.2930}{{\tt arXiv:1203.2930}}.

\bibitem{lusztig1984characters}
G.~Lusztig, {\em Characters of reductive groups over a finite field}, vol.~107.
\newblock Princeton University Press, 1984.

\bibitem{braden2014quantizations}
T.~Braden, A.~Licata, N.~Proudfoot, and B.~Webster, {\it Quantizations of
  conical symplectic resolutions ii: category $\mathcal{O}$ and symplectic
  duality},  {\em arXiv preprint arXiv:1407.0964} (2014).

\bibitem{proudfoot2014poisson}
N.~Proudfoot and T.~Schedler, {\it Poisson-de rham homology of hypertoric
  varieties and nilpotent cones},  {\em arXiv preprint arXiv:1405.0743} (2014).

\bibitem{dimoftetalk}
T.~Dimofte, {\it {Symplectic Duality and Knot Homologies}},  {\em (talk at
  String-Math 2014)}.

\end{thebibliography}\endgroup
